\documentclass[apjl]{emulateapj}

\newcommand{\OIII}{\mbox{O\,\textsc{iii}}}
\newcommand{\OII}{\mbox{O\,\textsc{ii}}}

\newcommand{\NII}{\mbox{N\,\textsc{ii}}}

\newcommand{\SII}{\mbox{S\,\textsc{ii}}}

\newcommand{\kms}{km s$^{-1}$}
\newcommand{\Ha}{H$\alpha$}   
\newcommand{\Hb}{H$\beta$}  
\newcommand{\msigma}{M$_{\rm BH}$-$\sigma_*$}
\newcommand{\msun}{M$_{\odot}$}
\newcommand{\ergs}{erg s$^{-1}$} 
\usepackage[T1]{fontenc}
\usepackage{lmodern}
\usepackage{hyperref}
\usepackage{natbib}
\usepackage{tikz}
\usepackage{kotex}
\usepackage[symbol]{footmisc}
\usepackage[T1]{fontenc}
\usepackage{etoolbox}
\usepackage{lipsum}

\usetikzlibrary{positioning,fit,calc}
\usetikzlibrary{shapes,backgrounds}

 \shorttitle{A systematic search for hidden type 1 AGNs}
 \shortauthors{Eun et al.}

\begin{document}

\title{A SYSTEMATIC SEARCH FOR HIDDEN TYPE 1 AGNS  :  GAS KINEMATICS AND SCALING RELATIONS}
\author{Da-in Eun$^{1,2}$}
\author{Jong-Hak Woo$^{1,}$\altaffilmark{$\dagger$}}
\author{Hyun-Jin Bae$^{1,3}$}

\affil{$^{1}$Astronomy Program, Department of Physics and Astronomy, Seoul National University, Seoul 151-742, Republic of Korea;}
\affil{$^{2}$Department of Earth Science Education, Seoul National University, Seoul 151-742, Republic of Korea;} 
\affil{$^{3}$Department of Astronomy and Center for Galaxy Evolution Research, Yonsei University, Seoul 120-749, Republic of Korea} 
\altaffiltext{$\dagger$}{Author to whom any correspondence should be addressed: woo@astro.snu.ac.kr}

\begin{abstract}
We search type 1 AGNs among emission-line galaxies, that are typically classified as type 2 AGNs based on emission line flux ratios if a broad component in the \Ha\ line profile is not properly investigated. Using $\sim$24,000 type 2 AGNs at z $<$0.1 initially selected from Sloan Digital Sky Survey Data Release 7 by \cite{Bae+14}, we identify a sample of 611 type 1 AGNs based on the spectral fitting results and visual inspection.
These hidden type 1 AGNs have relatively low luminosity with a mean broad \Ha\ luminosity, log L$_{\rm H\alpha }$ $=$ 40.73$\pm$0.32 \ergs\, and low Eddington ratio with a mean log L$_{bol}$/L$_{\rm Edd}$ $=$ -2.04$\pm$0.34, while they do follow the black hole mass - stellar velocity dispersion relation defined by the inactive galaxies 
and the reverberation-mapped type 1 AGNs. We investigate ionized gas outflows based on the [\OIII] $\lambda$5007 kinematics, which show relatively high velocity dispersion and velocity shift, indicating that the line-of-sight velocity and velocity dispersion of 
the ionized gas in type 1 AGNs is on average larger than that of type 2 AGNs. 
\end{abstract}


\section{Introduction}
The spectral features of AGNs (active galactic nuclei) are observed in different ways since the nucleus 
can be obscured by an optically thick dust torus \citep{Antonucci+93, Urry+95}. 
According to the simplest AGN unification model, the observed physical properties show differences due to the orientation effect, which depends on the angle between the line-of-sight and the axis of the dust torus \citep{Antonucci+93}. If we directly observe the inner part of an AGN, a blue AGN continuum from an accretion disk and broad emission lines 
originated from the broad-line region (BLR) are present in the optical spectral range. In contrast, if the dust torus obscures the inner part of an AGN, we observe narrow emission lines originated from the photoionized narrow-line region (NLR), without the direct detection of a blue AGN continuum and broad emission lines. 

If the FWHM of the broad emission line, e.g., \Hb\ and \Ha, is larger than 1000 \kms, AGNs are typically classified as type 1 AGNs \citep[e.g.,][]{Seyfert43}, for which the kinematics of gas in the BLR can be used to trace the gravitational potential of the central supermassive black hole for determining black hole mass \citep[e.g.,][]{Blanford+82, Kaspi+00,Peterson+04,Bentz+13}. In addition, the relativistic effect can also produce gravitational redshift causing velocity shift of the broad emission lines, as the BLR gas has orbital velocity corresponding to a few percent of the speed of light \citep[e.g.,][]{Zheng+90,Kollatschny+03,Tremaine+14}. 

When AGN luminosity is relatively low, however, the emission from host galaxy dilutes AGN emission, consequently dominating in the observed spectra. Thus, either a blue AGN continuum or broad emission lines can be missed from detection if no careful analysis is performed. In particular, a weak broad \Ha\ emission line from the BLR can be easily undetected if the narrow \Ha\ emission originated from the NLR and/or star forming region is very strong. 
Consequently, these low luminosity AGNs may be mis-classified as type 2 AGNs or star forming galaxies instead of type 1 AGNs. Previous studies by \cite{Woo+14} and \cite{Oh+15}, for example, showed that by decomposing the broad and narrow components in the \Ha\ line profile, a hidden population of type 1 AGNs can be identified among emission line galaxies \citep[see also][]{Greene+05}. 
The hidden population of type 1 AGNs is important to constrain the AGN unification model and to characterize dust torus
since the empirical number ratio between type 1 and type 2 AGNs is a key to understand the orientation effect \citep{Willott00, Simpson05,Oh+15}
although it is difficult to overcome the different bias and systematic difference in selecting type 1 and type 2 AGNs. Searching for intermediate mass black holes, other studies examined the \Ha\ line profile to investigate the presence of a broad \Ha\ component and identified type 1 AGNs \citep{Reines+13, Baldassare+16}. 

Hidden type 1 AGNs are characterized with low luminosity and low Eddington ratio \citep[e.g.,][]{Woo+14, Oh+15}. Thus, the optical continuum mainly represents stellar population while narrow emission lines, e.g., [\OIII] $\lambda$5007, are similar to those of type 2 AGNs. 
Compared to typical type 1 AGNs, these hidden type 1 AGNs have advantage in studying their host galaxies and investigating the coevolution of black holes and galaxies by constraining the black hole mass scaling relations  
\citep{Ferrarese00,Gultekin+09, Kormendy&Ho2013}. For example, while it is very difficult to measure stellar velocity dispersion or
bulge mass of type 1 AGN host galaxies due to the high flux ratios between AGNs and stars \citep[e.g.,][]{Woo+06, Park+15} 
it is easy to measure the properties of host galaxies if the AGN continuum is weak \citep{Reines+15, Woo+15}.
Thus these AGNs can be utilized to investigate the black hole-galaxy scaling relations and their evolution \citep{Barth+05, Woo+13}.

In our pilot study \citep{Woo+14}, we searched hidden type 1 AGNs among emission line galaxies at z $<$0.05, by carefully investigating the presence of a broad component in the \Ha\ emission line profile, using a large sample of $\sim$24,000 type 2 AGNs based on Sloan Digital Sky Survey (SDSS) Data Release (DR) 7 \citep{Bae+14}. 
In this paper  we extend our previous work by enlarging the survey volume out to  z=0.1, and improving the detection scheme
for the broad component in \Ha. Based on the newly identified sample of type 1 AGNs with a broad \Ha\ component, we investigate the \msigma\ relation, the kinematics of ionized gas outflows, and the effect of gravitational redshift. 
We present sample selection and analysis methods in Section 2 and 3, respectively. In Section 4, we present the results on the \msigma\ relation. 
In Section 5, we investigate the kinematics of [\OIII] and the gravitational redshift of the broad \Ha. Discussion is given in Section 6 and summary and conclusions follow in Section 7. Throughout the paper, we use the cosmological parameters as $H_0 = 70$~km~s$^{-1}$~Mpc$^{-1}$, $\Omega_{\rm m} = 0.30$, and $\Omega_{\Lambda} = 0.70$.

\section{Sample Selection}

 We utilized the type 2 AGN catalog by \cite{Bae+14} \citep[see also][]{Woo+16, Woo+17}, which identified type 2 AGNs at 0.02$<$z$<$0.1, in order to search hidden type 1 AGNs. The sample is selected based on a couple of criteria: signal-to-noise ratio (S/N)$>$3 for the four emission lines, namely, \Hb, [\OIII] $\lambda5007$, \Ha, and [\NII] $\lambda6583$ from the MPA-JHU catalog for SDSS DR7 galaxies \citep{abazajian+09}.
Using the flux ratios of these four emission lines \citep{Baldwin+81} and the demarcation line for separating AGNs from star-forming galaxies \citep{Kauffmann+03},  23,517 type 2 AGNs (pure AGNs and composite objects)were identified with the amplitude-to-noise ratio \textgreater 5 for [\OIII] and \Ha. 
More details of the selection procedure can be found in \cite{Bae+14} and \cite{Woo+16}.

To identify hidden type 1 AGNs among type 2 AGNs, we first subtract stellar continuum by fitting SDSS spectra using the penalized pixel-fitting code (pPXF) \citep{Cappellari+04}, which find the best-fit stellar population model of the given galaxy spectrum. We use MILES simple stellar population models with solar metallicity \citep{sb06}. In addition to the optical emission lines, we mask the continuum around the \Ha\ region (6300 $\sim$ 6900 \AA) to prevent the stellar population model from fitting a potential broad component as a stellar continuum.
 
After subtracting the best-fit stellar population model, we investigate whether a broad \Ha\ component is present around the \Ha\ region by visually 
inspecting the residual spectra (i.e., emission-line spectra). In this process we initially checked that $\sim$1000 objects show a broad feature around the narrow \Ha\ and [\NII] lines. To confirm the presence of a broad \Ha\ component, we need a reliable emission line decomposition process since the \Ha\ region is complex due to the blending of the narrow \Ha\ and the [\NII] doublet as well as the broad \Ha\ component if present. Therefore, we use three different fitting approaches for the \Ha\ region as follows: (a) If the narrow emission lines do not show particular wing components and if there is no broad \Ha\ component, we use a single Gaussian model for each of the \Ha\ and two [\NII] lines. In case of the [\NII] doublet, we fix their centers and flux ratio at their theoretical value while we use the same line width; (b) If wing components are present in the narrow emission lines (\Ha\ and [\NII] doublet), we utilize a double Gaussian model (one for a core component and the other for a wing component) to account for outflow kinematics in the NLR as we used for typical type 2 AGNs \citep{Woo+16} (c) If a very broad \Ha\ component with FWHM  > 1000 \kms\ is present, we add a single Gaussian component or a Gauss-Hermite component depending on whether this component is symmetric or asymmetric. For the broad and narrow \Ha\ components, we use free parameters for the line centers, widths and amplitudes. 

For the total sample, we first apply model (a) and identify type 1 AGN candidates by visually inspecting whether the model-subtracted spectra show a significantly high residual around the \Ha\ region. Note that the residual can be present due to a very broad \Ha\ component or wing components of the narrow \Ha\ and [\NII] lines. For these type 1 AGN candidates, we then apply either model (b) or (c) depending on the line profile of the [\SII]\ doublet. 
If the [\SII]\ doublet shows wing components based on a single Gaussian fit and visual inspection, then we apply model (b) by including a wing component for each of \Ha\ and [\NII], while we use model (c) only if both model (a) and (b) do not provide an acceptable fit (i.e., a significant residual is present based on visual inspection). In this process, we conservatively identify type 1 AGNs only for the case with model (c) in order to avoid false detection of the broad \Ha\ component. In total, we obtained 611 type 1 AGNs with a broad \Ha\ lines with FWHM > 1000 \kms. Note that we use FWHM $=$ 1000 \kms as a lower limit of the broad \Ha\ line width as a conventional definiton of type 1 AGNs \citep{Vanden+06} since a \Ha\ line narrower than 1000\kms\ may not be originated from the BLR. In Figure 1, we present examples with the best-fit models in the \Ha\ region.

  \begin{figure*}
  	\includegraphics[width=1\textwidth]{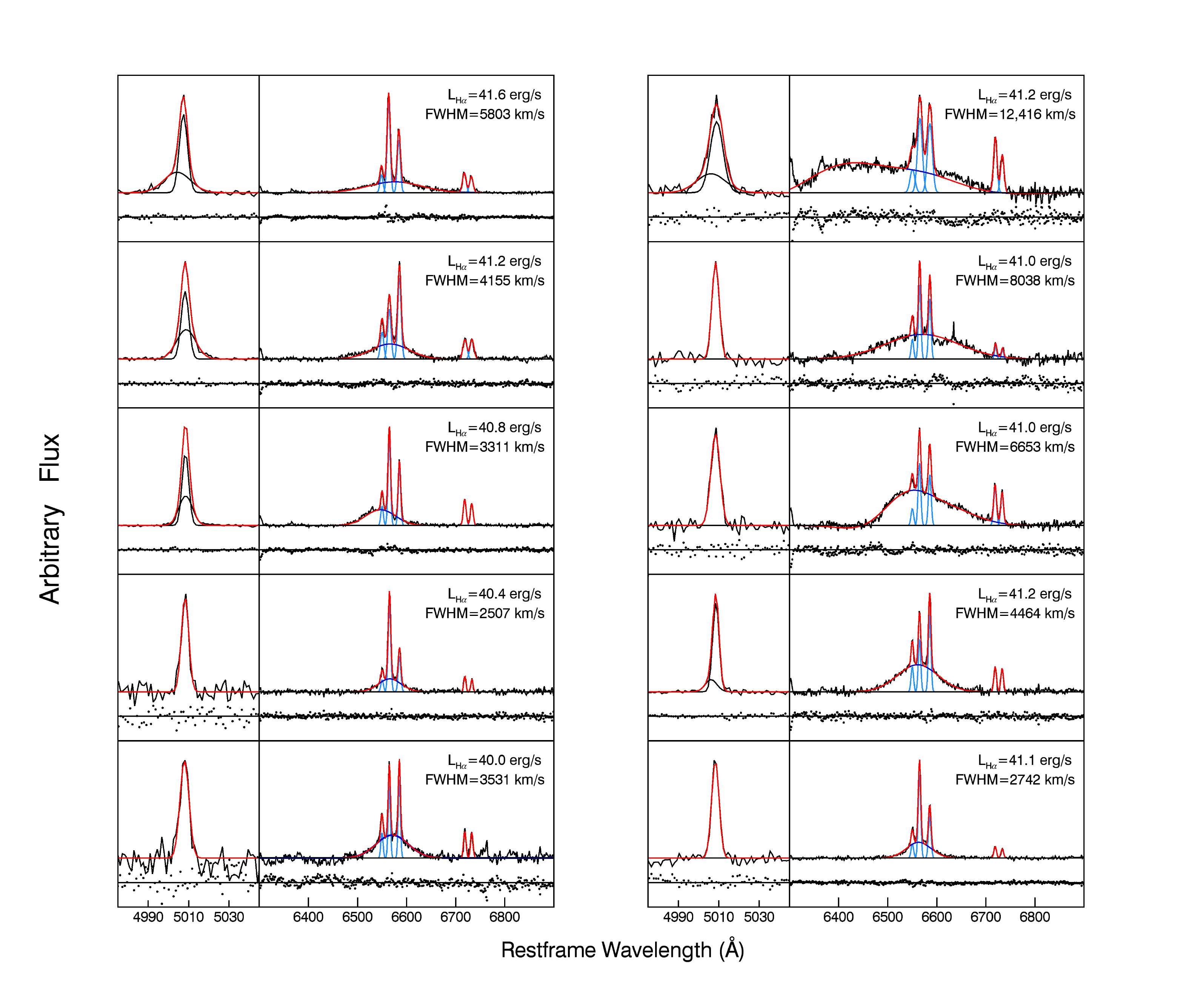}
  	\caption{Examples of the spectral decomposition of the \Ha\ region and the [\OIII] line. Continuum-subtracted spectra (black line) are overploted with the best-file model (red), which is composed of the narrow line models (light blue) and the broad component of \Ha\ (blue) in the right panels, while the [\OIII] line profile is modeled with either a single or double Gaussian profile in the left panels.
The residuals are presented at the bottom of each panel.
  	}
  \end{figure*}

\section{Analysis}
\subsection{Measurement of gas kinematics}

To investigate gas kinematics, we measure the velocity shift with respect to the systemic velocity based on stellar lines, velocity dispersion and luminosity of the [\OIII] $\lambda$5007. We fit [\OIII] using single or double Gaussian models to investigate the ionized gas outflows. First, we fit the total sample with a single Gaussian model. If the single Gaussian fit is not acceptable and a wing component is present based on visual inspection, we use a double Gaussian model. Since in some cases  a noise is fit with a second Gaussian component, we accept the double Gaussian fit only if the peak of the wing component is a factor of three larger than the noise level as we adopted in \citet{Woo+16}(see Figure 1).
Based on the best-fit model, we calculate the first and second moment of the total line profile as:
\begin{eqnarray}
\lambda_{0} = {\int \lambda f_\lambda d\lambda \over \int f_\lambda d\lambda}
\end{eqnarray}
\begin{eqnarray}
\sigma^2  = {\int \lambda^2 f_\lambda d\lambda \over \int f_\lambda d\lambda} - \lambda_0^2
\end{eqnarray}
where $f_{\lambda}$ is the flux density and $\sigma$ refers to the second moment line dispersion. 
For the given $\lambda_{0}$ , the velocity shift is calculated with respect to the systemic velocity. 
Note that we follow the same procedure to analyze the [\OIII] kinematics presented in our previous study of type 2 AGNs
\citep{Woo+16}.

\subsection{Morphology classification}

We use the Galaxy Zoo human classification data to obtain the host galaxy morphology \citep{Lintott+08,Lintott+11}. The host galaxies of the hidden type 1 AGNs are classified as 221 spirals ($\sim$36\%) and 381 ellipticals ($\sim$62\%). Nine ($\sim$1.5\%) of them are not classified due to the low image quality. 
Spiral host galaxies are further divided into face-on and edge-on galaxies based on the minor-to-major axis ratio (i.e., b$/$a) obtained from the SDSS DR7. If the ratio is larger than 0.5, we classify them as edge-on galaxies. Otherwise, we classify them as face-on galaxies. According to this criteria, 221 spirals are divided into 44 edge-on spirals and 177 face-on spirals. 

\subsection{Black Hole Mass and Eddington ratio}

To calculate black hole mass, we use the single-epoch virial mass estimator calibrated by \cite{Woo+15} as:
\begin{eqnarray}
M_{\rm BH}=f\times10^{6.819} \times \left ({\sigma_{\rm H\beta}  \over 10^{3} \rm\ km\;s^{-1}} \right)^{2} \times 
\nonumber\\
\left({\lambda L_{5100} \over 10^{44} \rm\ erg\;s^{-1} }\right)^{0.533}   M_{\odot}
\end{eqnarray}
where $\sigma_{\rm H\beta}$ is the line dispersion of the \Hb\ line and L$_{5100}$ represents  AGN continuum luminosity at 5100\AA.
However, since we do not detect AGN continuum and barely see the \Hb\ line due to their low flux in the large fraction of our sample, we use the width (i.e., FWHM and line dispersion) and the luminosity of the broad \Ha\ line as a proxy of \Hb. When we use the FWHM of \Ha\ broad component, we calculate the \Ha-based M$_{\rm BH}$ following \citet{Woo+15} as:
\begin{eqnarray}
M_{BH}=f \times 10^{6.544} \left ({L_{\rm H\alpha} \over 10^{42} \rm\ erg \;s^{-1}}\right)^{0.46}
\nonumber\\
\times \left ({\rm FWHM_{H\alpha} \over 10^{3} \rm\ km \;s^{-1}}\right)^{2.06} M_{\odot}
\end{eqnarray} 
where $L_{\rm H\alpha}$ is the luminosity of the broad \Ha\ and the FWHM$_{\rm H\alpha}$ is the width of the broad \Ha\ component.
When we instead use the line dispersion of the  broad \Ha\ ($\sigma_{H\alpha}$), we calculate M$_{\rm BH}$ following \citet{Woo+15} as: 
\begin{eqnarray}
M_{BH}=f \times 10^{6.561} \left ({L_{\rm H\alpha} \over 10^{42} \rm\ erg \;s^{-1}}\right)^{0.46}
\nonumber\\
\times \left ({\rm \sigma_{H\alpha} \over 10^{3} \rm\ km \;s^{-1}}\right)^{2.06} M_{\odot}
\end{eqnarray} 
We utilize log f=0.05 for the FWHM-based M$_{\rm BH}$ estimator and log f=0.65 for $\sigma$-based M$_{\rm BH}$, which were obtained from the recent calibration based on the \msigma\ relation of quiescent galaxies and \Hb\ reverberation-mapped AGNs \citep{Woo+15}. 
We use the luminosity of the broad \Ha\ component ($L_{\rm H\alpha }$) as a proxy for the bolometric luminosity ($L_{\rm bol}$), by utilizing the relation between \Ha\ and the continuum luminosity at 5100\AA\ \citep{Greene+05}, and the bolometric correction 9.26 for $L_{5100}$ \citep{Richards+06} as:
\begin{eqnarray}
L_{\rm bol}=2.21\times10^{44} ({L_{\rm H\alpha} \over 10^{42} \rm erg\;s^{-1}})^{0.86} \rm\ erg s^{-1}.
\end{eqnarray}

\section{Mass distribution and the scaling relations}

\subsection{Comparison with typical type 1 AGNs}

\begin{figure*}
	\center
	\includegraphics[width=.75\textwidth]{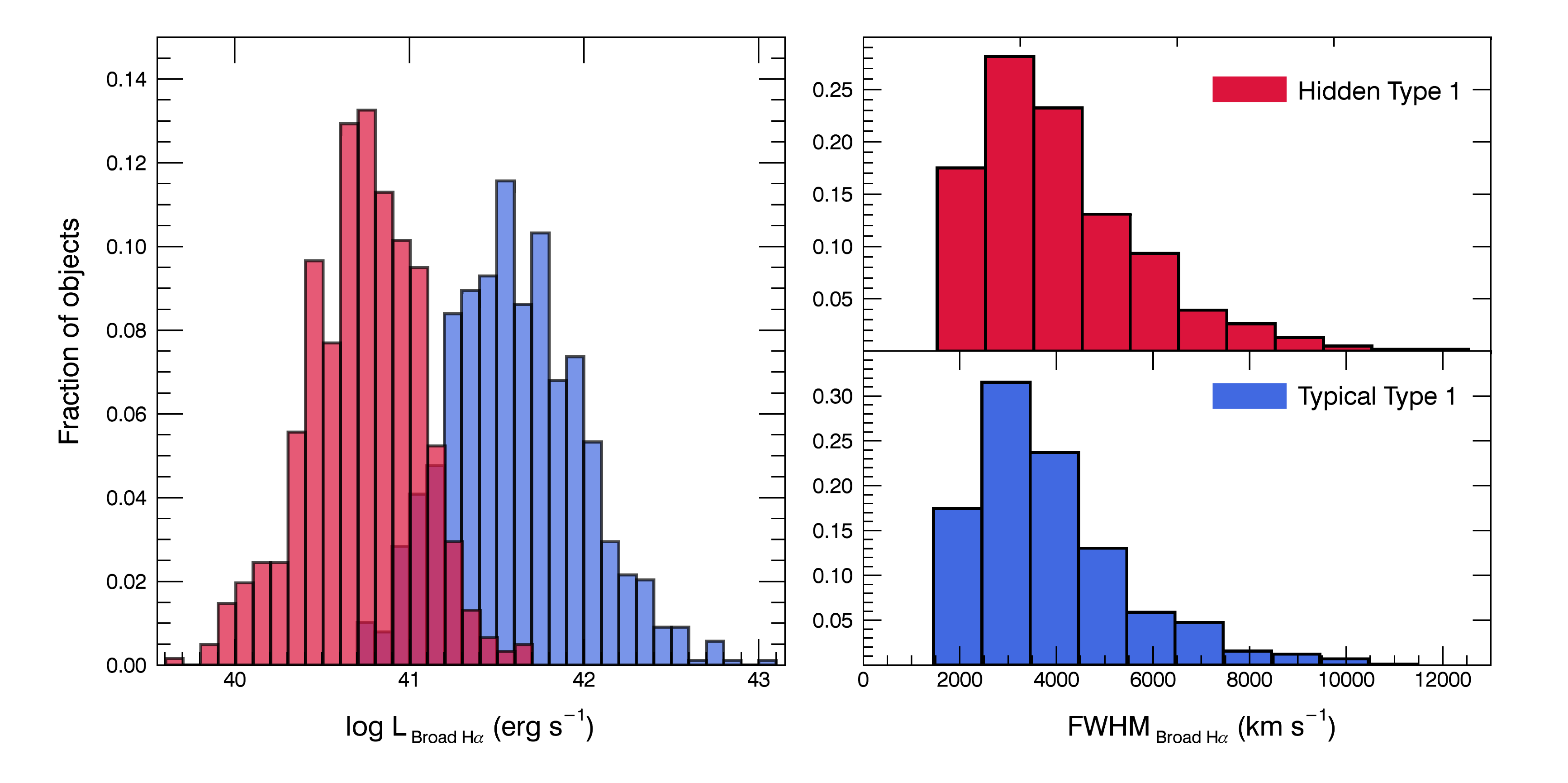}
	\caption{The normalized distributions of the \Ha\ luminosity (left) and FWHM (right), respectively for the hidden type 1 AGNs (red) and the comparison sample of type 1 AGNs (blue).}
\end{figure*}

\begin{figure}
	\center
	\includegraphics[width=.45\textwidth]{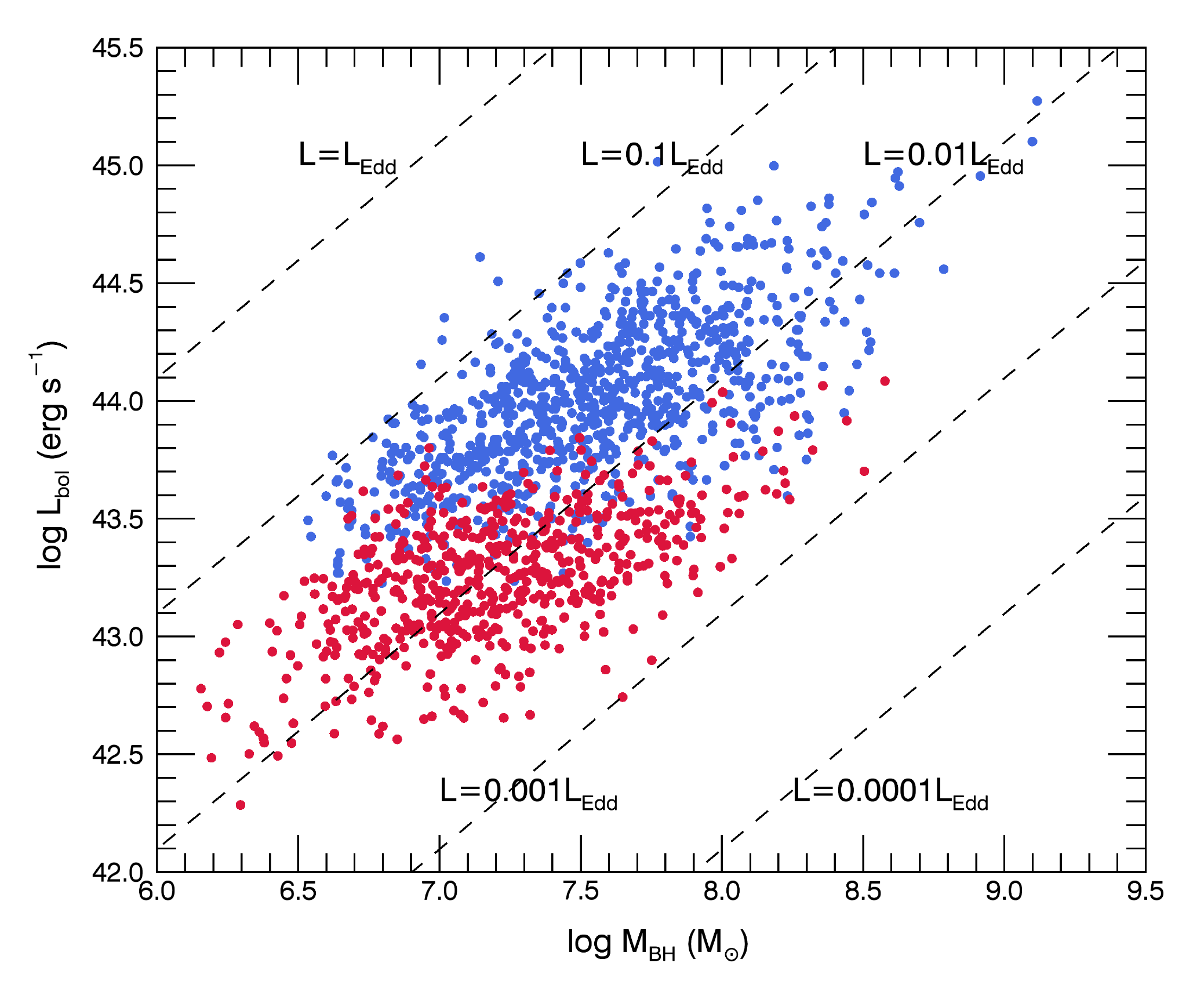}
	\caption{Black hole mass vs. bolometric luminosity of the hidden type 1 AGNs (red) and the comparison sample of type 1 AGNs (blue). Dashed lines represent the Eddington ratios.
	}
\end{figure}

In this section, we investigate the properties of the newly identified type 1 AGNs. The most prominent feature of these type 1 AGNs is that they have very weak AGN continuum, hence, the observed continuum represents the stellar continuum.  
To compare the hidden type 1 AGNs with typical type 1 AGNs, we choose 882 type 1 AGNs at the matched redshift range z$<$0.1
from the catalogue by \cite{Oh+15}. Using the luminosity and FWHM of the broad \Ha\ emission lines of the typical type 1 AGNs, we obtain bolometric luminosity, black hole mass and the Eddington ratio in the same way as we described in Section 3.3. Note that we use FWHM$_{\rm H\alpha}$ based M$_{\rm BH}$ in this section.

The hidden type 1 AGNs have relatively low broad \Ha\ luminosity while they have similar FWHM of the broad \Ha\ compared with typical type 1 AGNs (Figure 2). The mean \Ha\ luminosity of the hidden type 1 AGNs, $L_{\rm H\alpha }$, is $10^{40.73 \pm 0.32}$ \ergs\ while it is $10^{41.61 \pm 0.38}$ \ergs\ for the typical type 1 AGNs.
In contrast, there is no significant difference in the distribution of FWHM of \Ha\ (Figure 2 right panel).
Figure 3 presents the distribution of the hidden type 1 AGNs (red dots) and the typical type 1 AGNs (blue dots) in the $L_{\rm bol}$ - $M_{\rm BH}$ plane. The mean bolometric luminosity of the hidden type 1 AGNs is $10^{43.25 \pm 0.28}$ \ergs\ while that of typical type 1 AGNs is $10^{44.01 \pm 0.33}$ \ergs, indicating that the hidden type 1 AGNs have lower bolometric luminosity than the typical type 1 AGNs. In the case of black hole mass, the mean mass is $10^{7.20 \pm 0.42}$\msun\ and $10^{7.56 \pm 0.44}$\msun,
respectively for the hidden type 1 AGNs and the typical type 1 AGNs.
Combining mass and luminosity, the mean Eddington ratio of the hidden type 1 AGNs (i.e., 0.01 $\pm$ 0.01) is slightly lower than that of typical type 1 AGNs (i.e., 0.03 $\pm$ 0.02). 
Overall, the newly identified type 1 AGNs have relatively low luminosity while the width of \Ha\ is comparable to that of typical type 1 AGNs. 

\subsection{\msigma\ relation of the hidden type 1 AGNs}

\begin{figure*}
	\center
	\includegraphics[width=.95\textwidth]{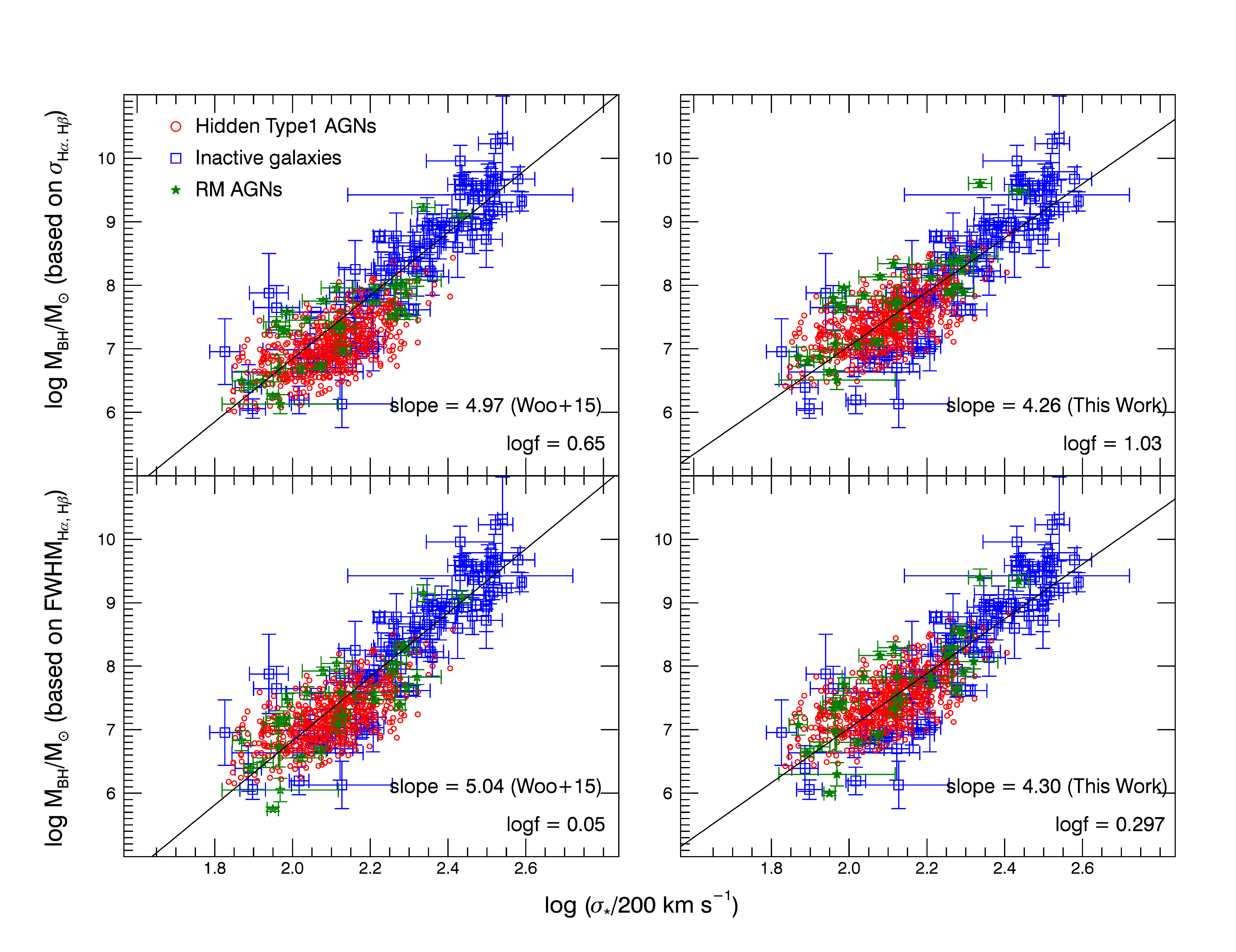}
	\caption{The hidden type 1 AGNs (red circle) are overploted with the \msigma\ relation calibrated based on inactive galaxies (blue square) and the reverberation-mapped AGNs (green star) by \cite{Woo+15} (left panels). 
We also present the joint-fit analysis of the \msigma\ relation (right panels) using the combined sample of the hidden type 1 AGNs,
inactive galaxies, and the reverberation-mapped AGNs based on Eq. 7. 
	}
\end{figure*}

We investigate whether the newly identified type 1 AGNs follow the black hole mass scaling relations as inactive galaxies and reverberation-mapped type 1 AGNs, using stellar velocity dispersion ($\sigma_*$) measurements from the SDSS DR7.
We adopt the most recent calibration by \cite{Woo+15} for $\rm \sigma_{\rm H\beta }$-based M$_{\rm BH}$ estimator as log (M$_{\rm BH}/M_{\odot})$$=$8.34 + 4.97 log ($\sigma_{*}$/200 \kms) and for FWHM$_{\rm H\beta}$-based M$_{\rm BH}$ estimator
as log (M$_{\rm BH}/M_{\odot})$ $=$ 8.34 + 5.04 log ($\sigma_{*}$/200 \kms). We find that the hidden type 1 AGNs locate slightly below the \msigma\ relation (Figure 4 left panels) with an average offset of 0.28 $\pm$ 0.40 dex and 0.12 $\pm $ 0.41 dex, respectively for the $\rm \sigma_{H\beta }$ and $\rm FWHM_{H\beta }$ based $M_{\rm BH}$ estimator.

The offset may be interpreted in different ways. First these hidden type 1 AGNs may not follow the same \msigma\ relation compared to typical type 1 AGNs. Second, since we use the \Ha\ velocity and luminosity as a proxy for \Hb\ velocity and 5100\AA\ luminosity in estimating black hole masses, systematic effects in the calibration of mass estimators may cause the offset, particularly at low mass scales where our targets are mainly located in the \msigma\ plane. Note that the empirical correlation between $L_{5100}$ and $L_{\rm H\alpha }$ has not been directly calibrated for the hidden type 1 AGNs. Third, there is a possibility that the virial factor may not be same for the hidden type 1 AGNs. Four, stellar velocity dispersion may be overestimated particularly for the late type galaxies since the significant rotational velocity may broaden stellar lines observed through the 3" aperture of the SDSS spectroscopy \citep[see e.g.,][]{Kang+13, Woo+13, Woo+15}. Without more detailed data, e.g., dynamical black hole mass measurements and spatially resolved kinematics measurements, it is difficult to evaluate these different scenarios. 

Instead, we further investigate whether the hidden type 1 AGNs have a different \msigma\ relation compared to inactive and reverberation-mapped AGNs 
by following the joint-fit analysis presented by \cite{Woo+15}. By combining the compilation of black hole mass and stellar velocity dispersion of inactive galaxies and reverberation-mapped AGNs from \cite{Woo+15} with those of our type 1 AGNs, we fit the \msigma\ relation log (M$_{BH}/M_{\odot})$$=$$\alpha $ + $\beta$ log ($\sigma_{*}$/200 \kms), in order to determine the virial factor, intrinsic scatter, intercept($\alpha $) and slope($\beta $) based on the $\chi^{2}$ minimization as
\begin{eqnarray}
{\chi^{2}}  = {\sum_{i=1}^{N}} {{{(\mu_{i} - \alpha -\beta s_{i})}^{2}} \over {{\sigma_{\mu , i}}^{2} +
		\beta^{2}{\sigma_{s, i}}^{2} + {\epsilon_{0}}^{2}}} + 
\nonumber\\
{\sum_{j=1}^{M}} {{{\rm (\mu_{VP,j} +\log f  - \alpha -\beta s_{j})}^{2}} \over {\rm {\sigma^{2}_{\mu _{VP,j}}} + \beta^{2}{\rm \sigma_{s, j}}^{2} + {\epsilon_{0}}^{2}}}
\end{eqnarray}
where $\mu=\log (M_{\rm BH}/M_{\odot})$ of inactive galaxies, $\mu_{VP}$ is the virial product of AGNs, and s represents log ($\sigma_{*}$/200 \kms). 
For our type 1 AGNs, we calculate the virial product by diving M$_{\rm BH}$ by the virial factor f in Eqs. 4 and 5.
As a result of joint-fit analysis, we obtained a slope of 4.26 and log$f$ $=$ 1.03 for the $\sigma_{\rm H\alpha }$-based M$_{\rm BH}$ estimator and a slope of 4.30 and log$f$ $=$ 0.30 for the $\rm FWHM_{H\alpha }$-based M$_{\rm BH}$ estimator (Figure 4 right panels). 
The offset of the hidden type 1 AGNs from the \msigma\ relation becomes negligible (i.e., 0.01 and 0.03 dex, respectively for $\sigma_{\rm H\alpha }$-based M$_{\rm BH}$ and FWHM$_{H\alpha }$-based M$_{\rm BH}$), and the scatter of the sample (0.37 dex) is comparable to
that of inactive galaxies and reverberation-mapped AGNs.
Based on the joint fit analysis we find no strong evidence that the hidden type 1 AGNs have a different \msigma\ relation compared to
inactive galaxies or the reverberation-mapped AGNs.

We examine whether the measured stellar velocity dispersion may be affected by the rotational velocity by comparing face-on and edge-on late type galaxies in the \msigma\ plane (see Figure 5).  As expected, we find that the majority of edge-on galaxies are located below the \msigma\ relation, suggesting that stellar velocity dispersion was overestimated due to the contribution of the rotational velocity, while face-on galaxies are more randomly distributed since the line-of-sight rotational velocity becomes much lower than that of edge-on galaxies.

In Figure 6, we compare the b/a ratio as a proxy for the inclination angle with the offset from the \msigma\ relation using late-type galaxies. 
We find a weak trend that the average offset increases with increasing b/a ratio, indicating that more inclined late type galaxies have larger
offset from the \msigma\ relation. These results suggest that spatially resolved kinematics are required to overcome the inclination and rotation effect
and to better investigate the \msigma\ relation of the hidden type 1 AGNs. Nevertheless, we find that the hidden type 1 AGNs follow the same \msigma\ relation as inactive galaxies and reverberation-mapped AGNs within the limitation of the current data.

\begin{figure}
	
	\includegraphics[width=.48\textwidth]{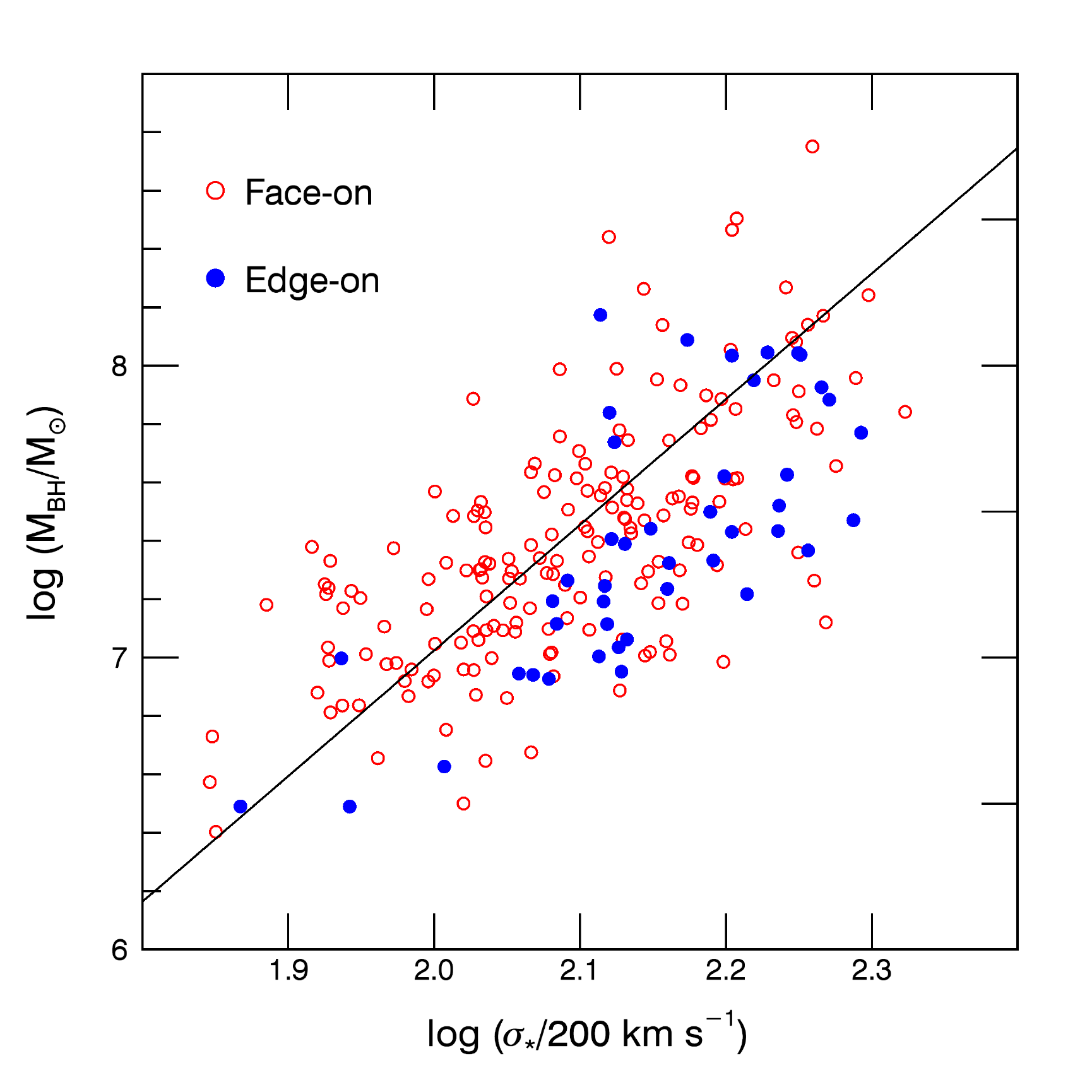}
	\caption{The \msigma\ relation of the hidden type 1 AGNs hosted by face-on late-type galaxies (red open circles) and edge-on late-type galaxies (filled blue circles). Black hole mass is based on $\rm FWHM_{H\alpha }$.
	}
\end{figure}

\begin{figure}
	\center
	\includegraphics[width=.48\textwidth]{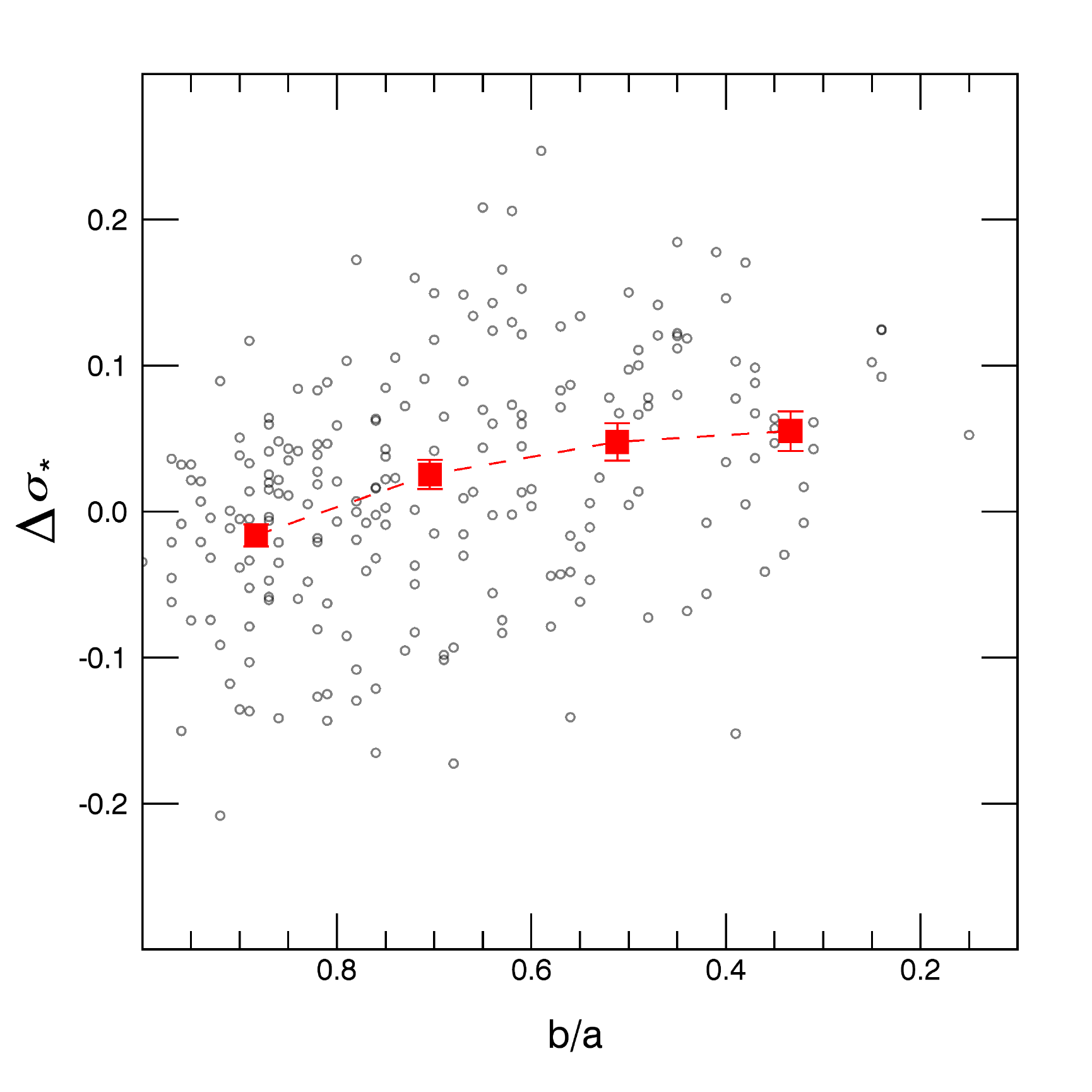}
	\caption{The offset from the \msigma\ relation in stellar velocity dispersion direction ($\Delta {\sigma_{*}}$) as a function of the b/a ratio. The mean $\Delta {\sigma_{*}}$ with a standard error is represented by red squares in each bin. 
	}
\end{figure}

\section{Gas kinematics}
\subsection{Narrow Line Region gas kinematics}

In this section, we investigate the gas kinematics in the NLR using [\OIII] and [\NII] emission lines compared to stellar kinematics. 
If the kinematics of NLR gas is governed by the gravitational potential, the velocity dispersions of emission lines and stellar absorption lines should be comparable  \citep{Komossa+08}. However, the ionized gas, e.g., [\OIII] of AGNs typically shows larger velocity dispersion than stars \citep{Woo+16, Woo+17}, suggesting the presence of non-gravitational component. 

First, we compare the velocity dispersion of [\NII] with stellar velocity dispersion, finding one-to-one relationship with a mean ratio, 0.99$\pm$0.20.
This suggests that [\NII] gas does not show any non-gravitational component and the broadening of [\NII] gas can be entirely explained by 
the gravitational potential. In the case of [\OIII], however, we find that the mean ratio between gas and stellar velocity dispersion is 1.22$\pm$0.50, indicating the presence of outflow component in [\OIII]. When we separately calculate the velocity dispersion of narrow core and wing components, the core component shows a consistent velocity dispersion compared to stellar velocity dispersion with a mean ratio 0.96$\pm$0.29 while the velocity dispersion of  the wing component of [\OIII] is a factor of 2.11$\pm$0.80 higher than stellar velocity dispersion, representing non-gravitational kinematics, i.e., outflows \citep[e.g.,][]{Boroson+05,Crenshaw+10, Woo+16, Woo+17}.

Second, we investigate the velocity shift of each emission line with respect to the systemic velocity (Figure 7). 
When we use the total profile of [\OIII], the mean velocity shift is -23$\pm$ 44 \kms\ with a skewed distribution toward blueshift. 
Compared to the distribution of [\OIII] velocity shift of type 2 AGNs \citep{Woo+16}, type 1 AGNs show a strong asymmetric distribution
with a much larger number of AGNs with blueshifted [\OIII] than AGNs with redshifted [\OIII], presumably due to the fact that
the direction of outflows is close to the line-of-sight (see the discussion on the number ratio between blueshifted and redshifted [\OIII] in \cite{Bae+16}).
Once we use the narrow core and broad wing components separately, the mean velocity shift is -10$\pm$38 \kms and -102$\pm$108 \kms, respectively. These measurements indicate that the core component of [\OIII] reflects the gravitational component while the wing component represents the outflow kinematics as discussed in the previous studies \citep[e.g.,][]{Boroson+05,Bae+14,Woo+16}.
We also measure the velocity shift of narrow \Ha, finding no significant velocity shift with a mean velocity shift 1$\pm$ 22 \kms\ and 
a similar distribution compared to the [\OIII] core component. 
In comparing velocity shifts of [\OIII] and \Ha, we find that the high ionization line is mainly blueshifted, which is consistent with the previous studies of type 1 AGNs \citep[e.g.,][]{Boroson+05,Komossa+08,Zamanov+02}. For instance, \cite{Boroson+05} reported that 77\% of the type 1 QSOs have blueshifted [\OIII], while low ionization lines (e.g., [\SII], [\NII], [\OII]) do not show any significant velocity shift.
Based on these results, we will use the wing component in [\OIII] as a tracer of gas outflows to further investigate AGN-driven gas outflow in the next section.

\begin{figure*}
	\center
	\includegraphics[width=.95\textwidth]{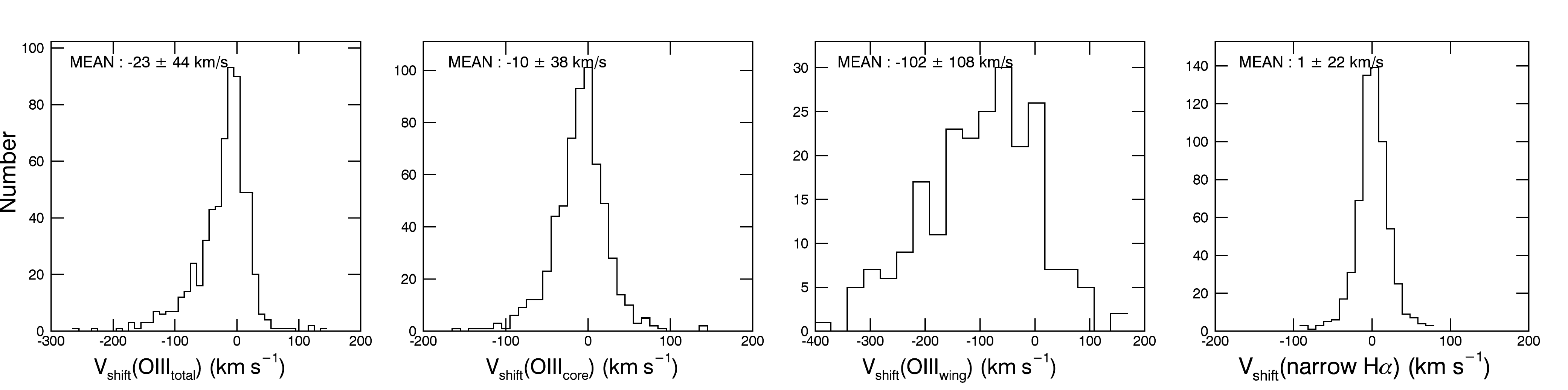}
	\caption{The distribution of the velocity shift of the [\OIII], respectively based on the total (left), core (center left), and wing profile (center right). The distribution of narrow \Ha\ velocity shift is also presented in the right panel. 
	}
\end{figure*}

\subsection{AGN-driven outflow}

\subsubsection{Velocity shift and velocity dispersion of [\OIII]}
Along with [\OIII] line dispersion, [\OIII] velocity shift is also an outflow signature \citep{Woo+16}. As presented in the Figure 7, the velocity shift of the [\OIII] total profile shows a blueward tail in its distribution. To estimate the velocity shift error in [\OIII] of each galaxy, we performed Monte Carlo simulation to produce 100 mock spectra. We measure the velocity shift in each mock spectra and consider the 1$\sigma$ dispersion as the uncertainty of velocity shift. The mean measurement uncertainty of the [\OIII] velocity shift is 17.0 $\pm\ $ 27.2 \kms. If we restrict the samples whose [\OIII] velocity shifts are larger than 1$\sigma$ measurement uncertainty, 414 AGNs ($\sim$ 68\%) show velocity shifts. Among the 414 AGNs, 318 AGNs ($\sim$ 77\%) are blueshifted. In type 2 AGNs, the blueshifted fraction is 56\%, when adopting the same measurement uncertainty using Monte Carlo Simulation \citep{Woo+16}. This result indicates that the blueshift is more common in type 1 AGNs than the type 2 AGNs \citep[see also][]{Crenshaw+10}. 
The relatively common blueshifts in type 1 AGNs can be explained by the combined model of biconical outflows and dust extinction \citep{Crenshaw+10, Bae+16}. When we observe type 1 AGNs, it is more likely that the receding cone is obscured by the dusty stellar disk, resulting in blueshift of the observed [\OIII] line profile. Therefore, [\OIII] of type 1 AGNs are more likely to be blueshifted than type 2 AGNs due to the orientation effect.

To further investigate kinematics of the AGNs with gas outflow in the NLR, we separate [\OIII] with a wing component (i.e., fitted with double Gaussian models; Group A) from [\OIII] without a wing component (i.e., fitted with single Gaussian; Group B).
Group A consists of 226 AGNs (about 34.2 \% of total sample) while Group B is composed of 385 AGNs. The mean [\OIII] velocity shift 
is -41 $\pm$ 54 \kms\ and -12.1 $\pm$ 31.5 \kms, respectively for Group A and Group B. Approximately 77\% of AGNs in Group A shows velocity shift larger than 1$\sigma$ measurement uncertainty, while  63\% of AGNs in Group B has velocity shift with >1$\sigma$ measurement uncertainty. If we count only these measurements (i.e., $>$ 1$\sigma$ error), the mean [\OIII] velocity shift is -49.5 $\pm$ 56.6 \kms\ and -18.9 $\pm$ 37.7 \kms, respectively for Group A and B, indicating that Group A has on average larger velocity shift than Group B (see Figure 8). 

In Figure 8 we compare the velocity shift and velocity dispersion of [\OIII]. The normalized [\OIII] line dispersions (log $\sigma_{\rm OIII}/\sigma_{*}$) of Group B are centered around 0, indicating that [\OIII] line dispersion and stellar velocity dispersion are comparable.
In contrast, Group A shows on average larger velocity shift and velocity dispersion than Group B. Compared to stellar velocity dispersion, [\OIII] 
velocity dispersion is much larger, indicating the presence of non-gravitational kinematics, i.e., outflows.
There is a trend that velocity shift becomes larger with increasing velocity dispersion as similarly found in type 2 AGNs \citep{Woo+16}, which 
is a characteristic feature of biconical outflows as demonstrated by \cite{Bae+16}.

\begin{figure}
\center
\includegraphics[width=.48\textwidth]{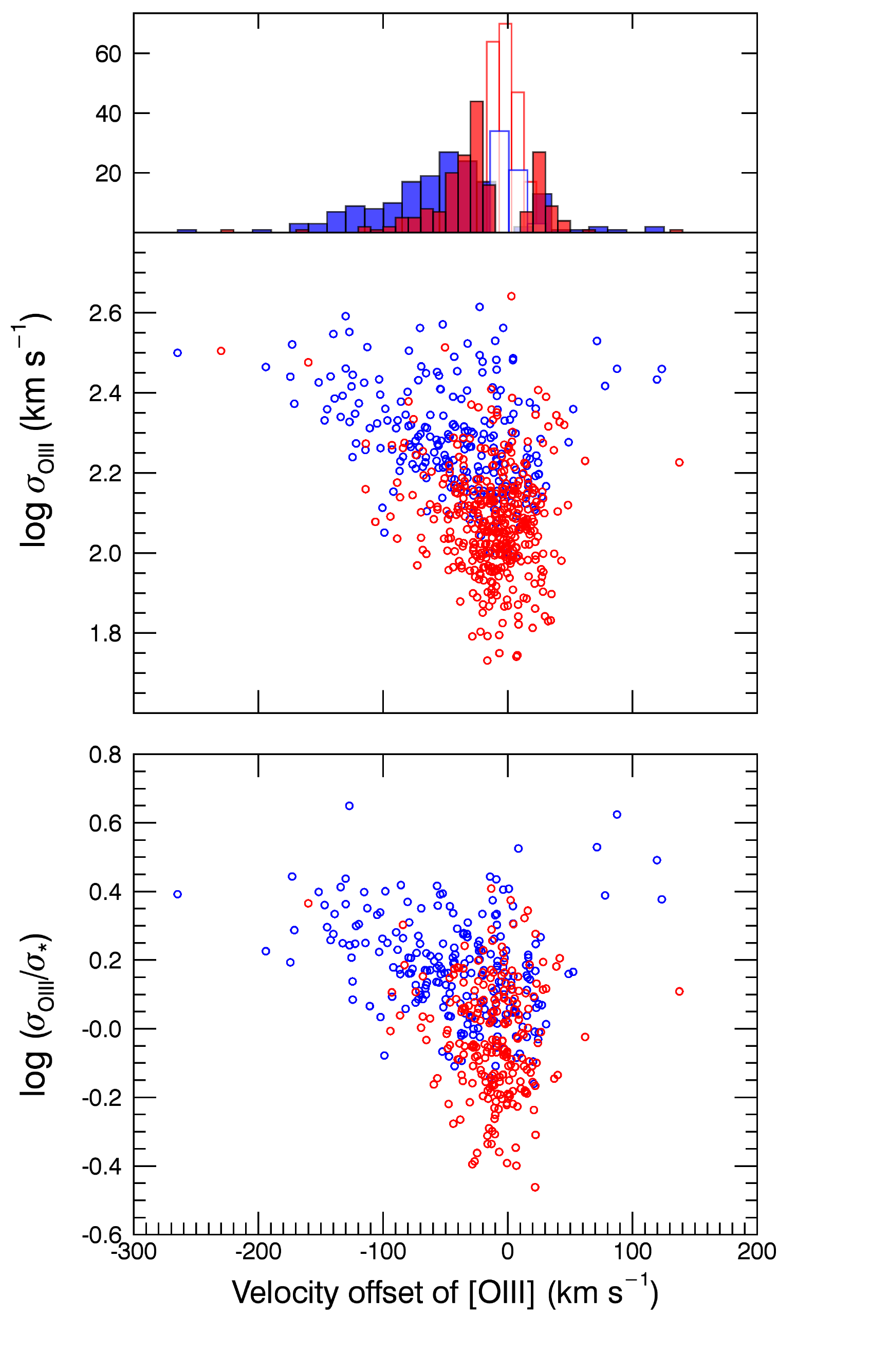}
\caption{Top: The distribution of the velocity shift of Group A (blue) and Group B (red) with reliable measurements with $<$1$\sigma$ measurement uncertainty (filled histogram) or the total sample (open histogram). Middle: [\OIII] velocity dispersion and velocity shift of Group A (blue circles) and Group B (red circles).  Bottom: The velocity shift is also compared with the normalized [\OIII] dispersion by stellar velocity dispersion.
}
\end{figure}

\subsubsection{Outflow fraction}

\begin{figure}
\center
\includegraphics[width=.42\textwidth]{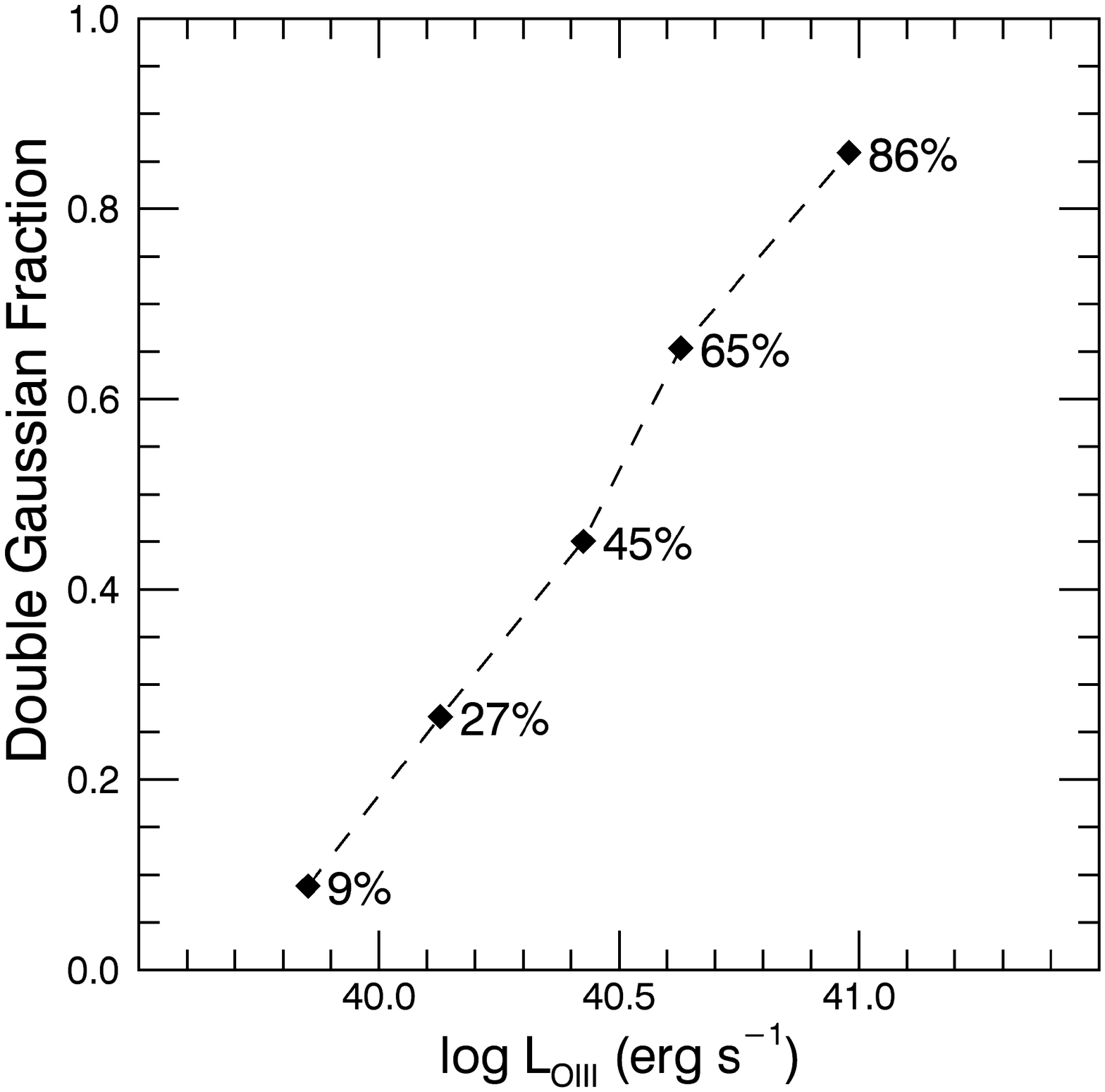}
\caption{The fraction of AGNs with a double Gaussian [\OIII] as a function of [\OIII] luminosity. 
}
\end{figure}

We investigate the outflow fraction as a function of [\OIII] luminosity and the Eddington ratio. 
Since the wing component of [\OIII] indicates gas outflows, we use the presence of the wing in [\OIII] as an evidence of outflows. 
The outflow fraction (i.e., the fraction of double Gaussian [\OIII]) rapidly increases with increasing luminosity (Figure 9). At high luminosity, e.g., $L_{[\OIII] }$ > $10^{41.0}$ \ergs, the outflow fraction is over 85\%, suggesting that most high-luminosity AGNs
have strong outflows, which is consistent with the trend found in type 2 AGNs \citep{Woo+16}. 
On the other hand, the outflow fraction in low luminosity AGNs is lower than that of type 2 AGNs. 
For example, the outflow fraction of AGNs at $10^{40.0}$ \ergs\ is 40\% in the case of type 2 AGNs, while it is $\sim$20\% in the hidden type 1 AGNs. It is not clear what is the origin of this discrepancy. A systematic comparison of the outflow fraction between type 1 and type 2 AGNs
is yet to be available. The detection of a wing component in [\OIII] is more difficult for lower luminosity AGNs since the wing component can be easily diluted by noise. 
Thus, the outflow fraction of the low luminosity AGNs should be regarded as a minimum \citep[see also][]{Woo+17}. 
We note that our hidden type 1 AGN sample is far from complete at low luminosity, since a broad \Ha\ component will be much weaker and easily missed in identifying type 1 AGNs. Therefore, we interpret the difference of the outflow fraction at low luminosity is insignificant.

The outflow fraction also increases as Eddington ratio increases albeit with a shallower slope. 
Over the Eddington ratio range covered by our sample, i.e., log $(L_{\rm bol}/L_{\rm Edd})$ = -2.5 to -1.5, the outflow fraction increases
from $\sim$30\% to $\sim$50\%. 
These results indicate that ionized gas outflows are directly connected to AGN activity and the outflow properties are qualitatively similar 
between type 1 and type 2 AGNs.

\subsection{Gravitational Redshift}

The velocity of BLR gas is relatively high up to several percent of the speed of light, suggesting that the the broad lines can be redshifted due to relativistic effects \citep[e.g.,][]{Kollatschny+03}. To investigate the gravitational redshift using the newly identified type 1 AGNs, we calculate the velocity shift of the broad \Ha\ with respect to the systemic velocity. In contrast to typical type 1 AGNs, for which systemic velocity cannot be accurately measured \citep{Tremaine+14}, we are able to use the luminosity-weighted stellar absorption lines to determine the systemic velocity since AGN continuum is too weak to dilute stellar absorption lines. Thus, we can decrease the systematic effect in measuring velocity shift of broad emission lines caused by the large uncertainty in systemic velocity.

In Figure 10, we present the velocity shift of the broad \Ha\ line as a function of the line width of \Ha. The mean velocity shift is 115 $\pm$ 389 \kms\ while 427 out of 611 AGNs show a redshifted \Ha\ ($\sim$ 70\%), which may be interpreted as the gravitational redshift
although the velocity shift may be also caused by other mechanisms such as non-gravitational kinematics (i.e., inflows and outflows) 
or orbital motions \citep{Zheng+90}. Note that the other 1/3 of the sample shows a blueshifted \Ha. We find several extreme objects with a very large blue shift ($<$ -1000 \kms) and a large velocity dispersion, while the \Ha\ line profile is very asymmetric. The nature of these objects is unclear and beyond the scope of the
current study.

Using the SDSS quasar sample, \cite{Tremaine+14} statistically investigated the velocity shift of \Hb\ to compare with the prediction of
their BLR geometry models. The [\OIII] narrow emission line was used to calculate the systemic velocity since stellar absorption lines are not detected due to strong AGN continuum. Comparing with their result (Figure 1 in \cite{Tremaine+14}), we obtain a similar result
within the observed range of the broad \Ha\ velocity dispersion $<$ $\sim$3000 \kms\ in our sample. 

The geometry of the BLR is often assumed with spherical or disk models, and these models predict that the effect of the gravitational redshift increases with the BLR gas velocity \citep{Tremaine+14}. By taking the expected velocity shift of the broad \Ha\ as 
 V$_{\rm H\alpha}$ = 1.5$\sigma_{\rm H\alpha}^{2}/ c$ based on a spherical model of BLR from \cite{Tremaine+14}, we compare the observed mean velocity shift with the prediction as a function of the line dispersion of broad \Ha. Although there is a large scatter of
 velocity shift for given velocity dispersion, the trend of the mean velocity shift is consistent with the model, suggesting that the observed
 velocity shift is consistent with the gravitational redshift effect.

\begin{figure}
\center
\includegraphics[width=.45\textwidth]{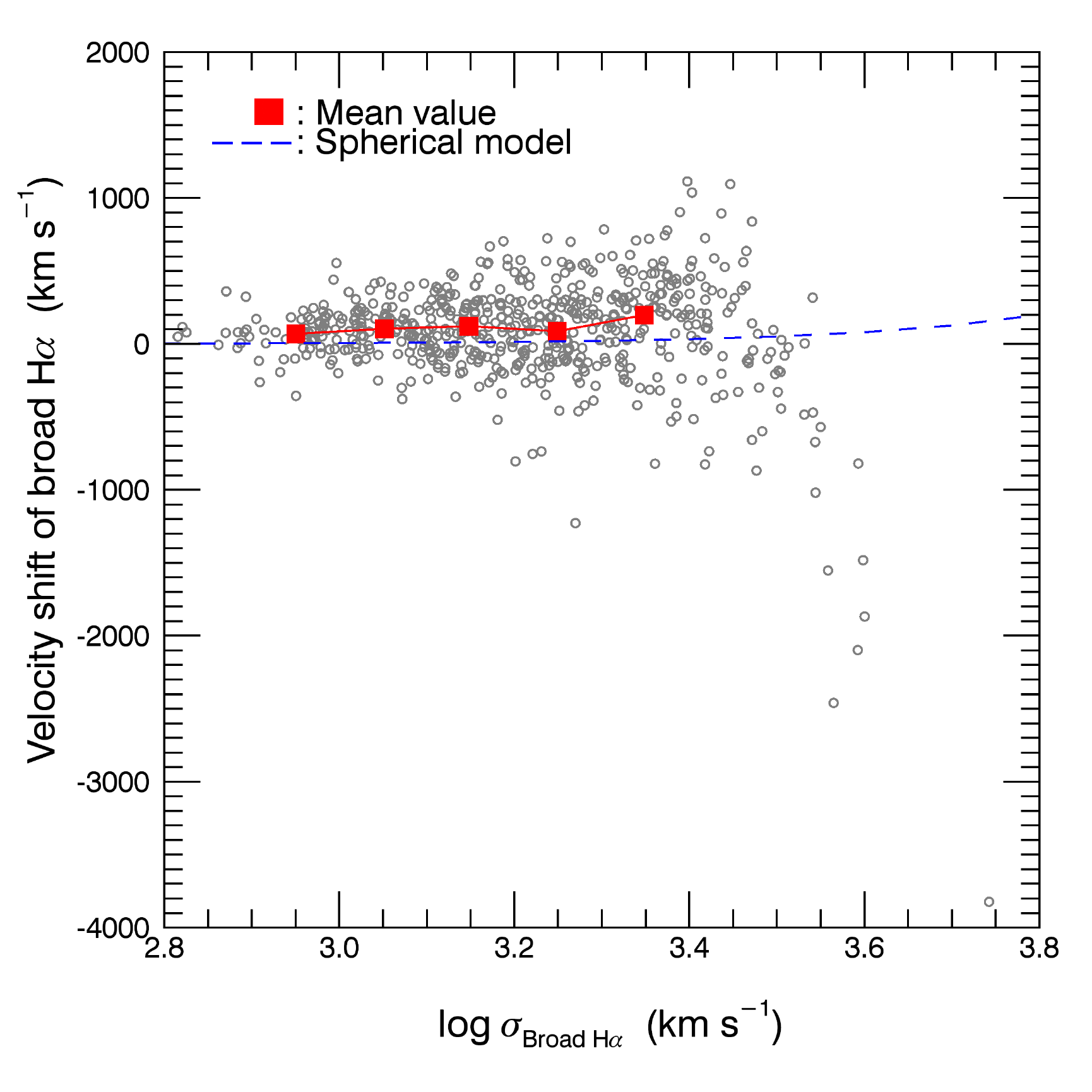}
\caption{Comparison of the velocity dispersion and velocity shift of the broad \Ha\ component. The red square symbols represent the mean value in each bin while the blue dashed line represents the expected velocity shift based on the spherical model of the BLR.
}
\end{figure}

\section{Discussion : Comparison with previous works}

\begin{figure}
	\center
	\includegraphics[width=.45\textwidth]{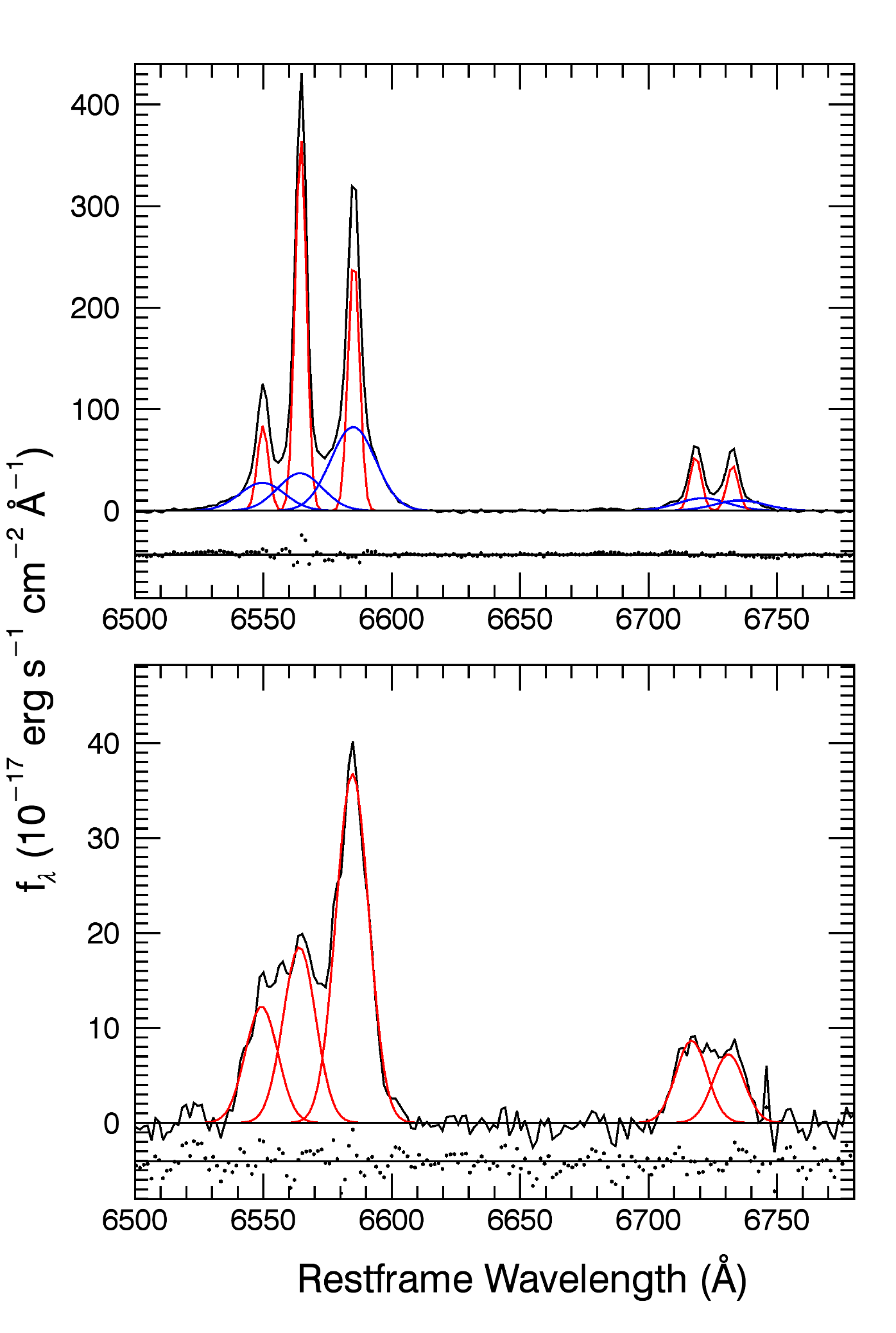}
	\caption{Examples of type 2 AGNs that were classified as type 1 AGNs by \cite{Oh+15}. As the [\SII] doublet clearly shows a wing feature (blue lines), double Gaussian models (blue + red) provide a reasonable fit for the \Ha\ region (top panel), demonstrating that a broad \Ha\ component is not required once we include a wing component in the narrow line profile. Bottom panel shows that for some objects, a single Gaussian model provide a good fit for all narrow lines (i.e., \Ha, [\NII], and [\SII]), 	
	}
\end{figure}

\begin{figure}
	\center
	\includegraphics[width=.49\textwidth]{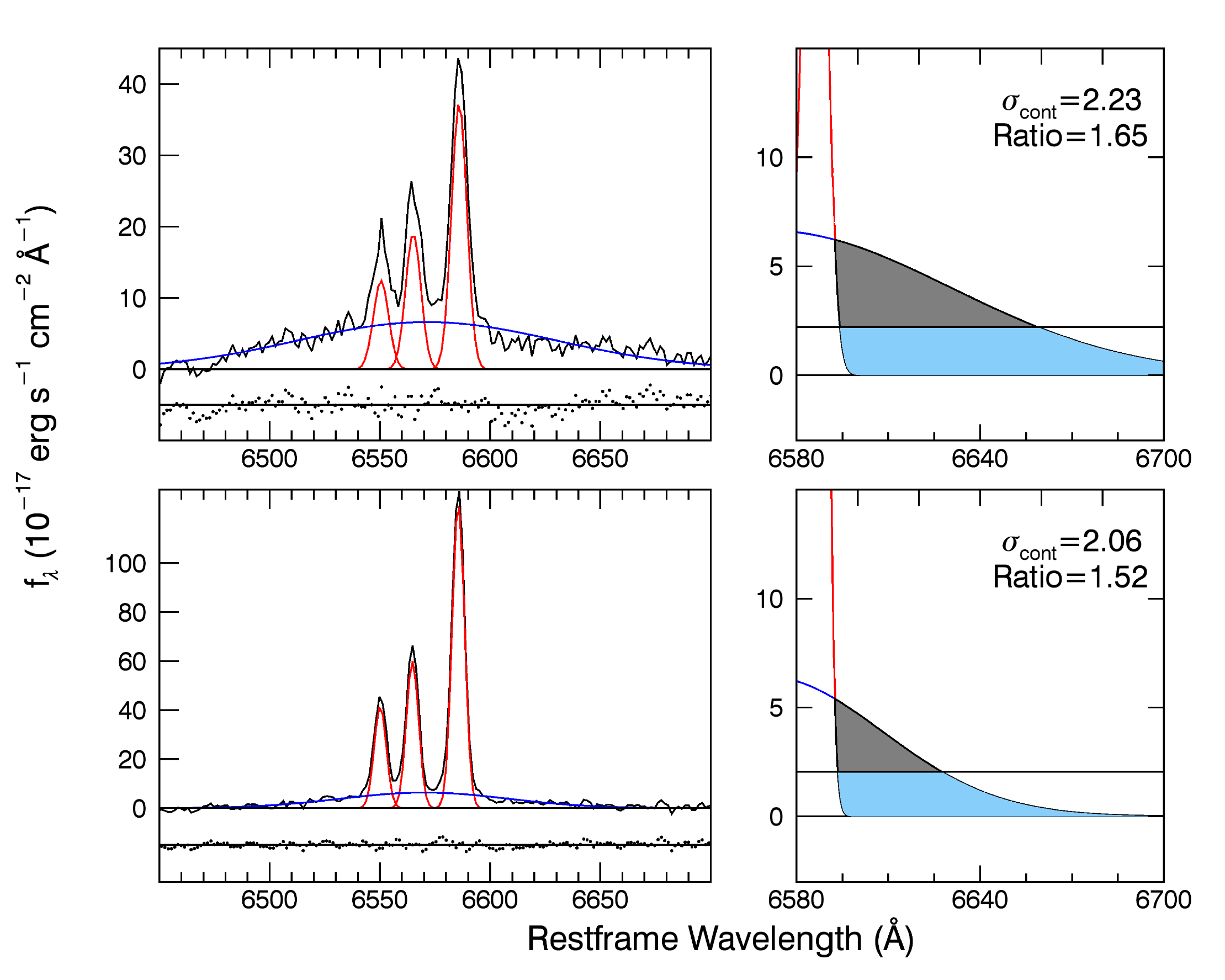}
	\caption{Examples of type 1 AGNs that were missed by \cite{Oh+15}  due to the area ratio criterion. 
While there is a broad \Ha\ component (blue line in the left panel), the area ratio is less than 2  since the noise area (blue section) is relatively
large due to the high noise level ($\sigma_{cont}$) (right panel). The area ratio is defined by the wing of the broad \Ha\ model (grey + blue sections) divided by the noise area (blue section), for which the height is determined base on the continuum noise level. 
	}
\end{figure}

Although the presence of broad emission lines is a characteristic of optical type 1 AGNs, in some cases it is difficult to detect 
broad lines due to obscuration in the optical spectral range (e.g., red AGNs) or intrinsically low luminosity diluted by host
galaxy emission \citep[e.g.,][]{Simpson05, Gilkman+07, Gilkman+13, Woo+14, Oh+15, Kim+15}. 
Several studies tried to find hidden type 1 AGNs using different methods.  For example, using the SDSS DR2 catalogue, \cite{Simpson05} found 63 objects (1.8\% of total galaxies), which have a relatively broad \Ha\ line based on visual inspection. Similar works by \cite{Woo+14}  and \cite{Oh+15} reported the detection of hidden type 1 AGNs using SDSS DR7. 

We compare the selection methods and results between our work and the previous studies. 
In our pilot study \citep{Woo+14}, we identified type 1 AGNs at 0.02$<$z$<$0.05 by detecting a broad component in the \Ha\ line profile. 
Using the same type 2 AGN catalogue, we extend our previous work out to z=0.1, enlarging the sample size from 142 to 611 objects. 
However, we note that the selection scheme is improved based on the revised spectral decomposition and visual inspection. For example, we model the \Ha\ region more carefully using 3 different cases as described in Section 2.2, while \cite{Woo+14} used a simple fitting scheme using one broad and one narrow components for \Ha. Thus, readers are suggested to use the results in this extended work. 

\cite{Oh+15} searched type 1 AGNs among emission line galaxies at z$<$0.2 using the SDSS DR7 catalogue. First, they used the flux ratio at two narrow band regions, respectively representing a broad \Ha\ (6523-6543\AA) and continuum (6460-6480\AA), to trace the presence of a broad \Ha\ component. This flux ratio $F_{6533}/F_{6470}$ is useful to find broad \Ha\ candidates since it increases as the broad \Ha\ line becomes more prominent. Combining the signal-to-statistical noise (defined from continuum spectral range) and the flux ratio $F_{6533}/F_{6470}$, they defined the 1$\sigma$ demarcation line to select type 1 AGN candidates \citep[see Figure 1 in][]{Oh+15}. 
Second, they used the area flux ratio, which is defined by the area of the red wing of the broad \Ha\ component, after excluding the [\NII] $ \lambda$6584 emission line (see the blue + grey area in Figure 12), divided by the noise area defined with the 1$ \sigma $ noise level of the continuum (blue area in Figure 12), in order to avoid unreliable detection of a broad \Ha. 
Only if the area flux ratio is larger than 2, then they classified a target as type 1 AGNs.
Using these criteria, they identified 1611 type 1 AGNs at z < 0.1.  Among these AGNs, 882 objects are SDSS specClass =3 objects, which are already classified as type 1 AGNs in the SDSS, while the other 729 objects are SDSS specClass =2 (i.e., galaxy) objects. Thus, they found 729 hidden type 1 AGNs among type 2 or star forming galaxies by detecting a broad \Ha, while in our study we found 611 type 1 AGNs from our type 2 AGN catalogue.

Among 729 type 1 AGNs identified by \cite{Oh+15}, 230 objects are overlapped with our sample of 611 type 1 AGNs, while the other 499 objects are not identified as type 1 AGNs in our study. When we examine these 499 objects, the majority of them (412 objects) are located in the star forming region in the BPT diagram. Since we only used the type 2 AGN catalogue to search hidden type 1 AGNs, we simply did not investigate these 412 AGNs. The remaining 87 objects were not classified as type 1 AGNs in our study since a broad \Ha\ component was not required to fit the \Ha\ region. Instead, we were able to fit \Ha\ and [\NII] with double Gaussian models for the majority of these objects (70 objects, see the top panel in Figure 11) or single Gaussian models provide a good fit for 7 objects (see the bottom panel in Figure 11). For some cases (10 objects), the noise in the vicinity of \Ha\ region is too large to decide whether single or double Gaussian models provide a better fit.

Note that in our fitting procedure we often used double Gaussian models whenever a wing component in the narrow line profile is required. In particular, since we also examined the line profile of the [\SII] doublet, in order to check whether a wing component is necessary to fit the narrow component of \Ha\ and [\NII] lines, we tried to avoid a false detection of fitting the wing components of narrow lines
as a broad \Ha\ component. In contrast, \cite{Oh+15} used single Gaussian models for narrow \Ha\ and [\NII] lines, hence they may have treated the wing components of the narrow lines as a broad \Ha\ component. In fact, the [\SII] line profile clearly shows the presence of a wing component for $\sim$a half of these objects (43/87) as demonstrated in Figure 11 (top panel). Since we used a double Gaussian profile for narrow lines (\Ha, [\NII], and [\SII]), our fitting method for detecting a broad component of \Ha\ seems to be more conservative than that of \cite{Oh+15}. 

We also investigate why 315 objects among 611 objects in our hidden type 1 AGN sample were not classified as type 1 AGNs by \cite{Oh+15} .
Many of them were excluded due to the area ratio criterion by \cite{Oh+15}. For example, when we apply the area ratio cut defined by \cite{Oh+15} to the 315 AGNs, 136 objects do not satisfy the criterion while in the current study we conservatively concluded that a broad \Ha\ is present (see Figure 12). In particular, if the noise level is relatively large, the area flux ratio becomes below 2, hence, these AGNs will be rejected by the criterion defined by \cite{Oh+15}. In Figure 12 we demonstrate two cases, each of which clearly shows a broad \Ha\ component in visual inspection while the area flux ratio does not satisfy the criterion (i.e., ratio$<$2). 
In addition, if the broad \Ha\ is blueshifted, the area ratio becomes also smaller since the area defined in the red side of the \Ha\ becomes smaller. 
For the remaining 179 objects among 315 objects, we find that the area flux ratio is slightly larger than 2 when we calculate it using our own noise estimates. Thus, it is not clear why these objects were rejected by \cite{Oh+15}. It is possible that there is difference in noise calculation and in modeling continuum between these two studies. 

In summary, although the selection method of \cite{Oh+15} provides consistent and quantitative criteria for identifying a hidden broad \Ha\ component, it is somewhat limited due to the fact that the wing components in the narrow lines are not included in their fitting, and that their area ratio cut may reject a weak and shallow broad \Ha\ component. More detailed studies on the detection scheme of a weak \Ha\ component is necessary to better detect hidden type 1 AGNs.

\section{SUMMARY AND CONCLUSIONS}

By detecting a broad component in the \Ha\ line profile, we conservatively identified a sample of 611 type 1 AGNs at 0.02 $<$ z $<$ 0.1, using the catalogue of \cite{Bae+14}, which provides a large sample of type 2 AGNs classified based on the emission line flux ratios. We increased the sample size of the hidden type 1 AGNs from 142 \citep{Woo+14} to 611 by extending the redshift range out to z=0.1 and using a consistent and improved emission line analysis. The main findings are summarized as follows.

$\bullet$
The hidden type 1 AGNs have a similar range of the \Ha\ width, but on average lower luminosity compared to typical type 1 AGNs at the same redshift range, indicating that these hidden type 1 AGNs with a relatively weak AGN continuum is useful to study the properties of low luminosity AGNs and their host galaxies. The mean black hole mass estimated based on the \Ha\ FWHM and luminosity is log M$_{\rm BH}$ =7.20$\pm$0.42, while the mean Eddington ratio is log L$_{\rm bol}$/L$_{\rm Edd}$ = -2.04$\pm$0.34. 	 

$\bullet$   
These AGNs seem to slightly offset from the \msigma\ relation defined by inactive galaxies and reverberation-mapped AGNs \citep{Woo+15}, presumably due to the systematic difference in estimating black hole mass (i.e., depending on \Hb\ or \Ha) and potentially different virial factors between typical type 1 AGNs and the hidden type 1 AGNs with a weak AGN continuum.  In contrast, when we perform a joint-fit analysis
by combining quiescent galaxies, RM AGNs, and hidden type 1 AGNs, we find no significant difference between
typical type 1 AGNs and hidden type 1 AGNs. 

$\bullet$                                                                                                                   
By investigating the kinematics of ionized gas, we find that the velocity dispersion of [\NII] and the core component of [\OIII] is roughly consistent with stellar velocity dispersion, indicating that the host galaxy gravitational potential is responsible for the broadening of these lines. In contrast, the wing component of [\OIII] represents non-gravitational kinematics, i.e., outflows, which are consistent with the finding of type 2 AGNs \citep{Woo+16, Woo+17}. 

$\bullet$
The velocity dispersion and velocity shift of [\OIII] show strong non-gravitational kinematics, i.e., outflows.
The fraction of AGNs with a wing component in [\OIII] strongly increases with AGN luminosity, suggesting that the non-gravitational
kinematics are directly connected to AGN activity. 

$\bullet$
The line-of-sight velocity and velocity dispersion of the ionized gas in type 1 AGNs is on average larger than that of type 2 AGNs \citep[e.g.][]{Woo+16}, which is consistent with the biconical outflow models and orientation effect \citep[e.g.,][]{Bae+16}.
Although the uncertainty of the velocity shift of [\OIII] is large, we find that there are more AGNs with
blueshifted [\OIII] than AGNs with redshifted [\OIII].

\medskip

Based on these results, we conclude that the hidden type 1 AGNs follow the general characteristics of typical broad line AGNs except for low luminosity and low Eddington ratio.
More detailed studies on the detection scheme and the completeness of the broad line AGNs may provide a better understanding
of these hidden type 1 AGNs, which can be used as a unique channel for studying AGN unification and black hole-galaxy connection. 

\acknowledgments
We thank the anonymous referee for valuable comments.
Support for this work was provided by the National Research Foundation of Korea grant funded by the Korea government (No. 2016R1A2B3011457 and No. 2010-0027910).

\bibliographystyle{apj}

\end{document}